\documentstyle[12pt]{article}
\begin {document}

\title{Some Methods of Minimization of Calculations in High Energy
Physics\footnote
{
Presented at the XVIII  International Workshop on High Energy
Physics and Field Theory, Protvino, Russia, June 26--30, 1995
} }
\author {Alexander L. Bondarev
\and \it National Scientific and Educational Center of Particle and
\and \it High Energy Physics attached to Belarusian State University
\and \it M.Bogdanovich str.,153, Minsk 220040, Republic of Belarus
\and \rm e-mail: bondarev@hep.by}

\setlength{\textheight}{218mm}
\setlength{\textwidth}{150mm}

\maketitle

\begin{abstract}

    Two approaches  to calculations'  minimization in  High Energy
Physics are considered.  The first one is the method of  covariant
calculations for the amplitudes of processes with polarized  Dirac
particles.  The second one connects with the possibility to reduce
the expressions for the traces  of products of ten and  more Dirac
$\gamma$-matrices.

\end{abstract}

\section {Introduction}

    It  is  well  known  that  the  high  order calculation of the
observables within  the perturbative  theory turns  to the serious
difficulties (especially, if we take into account the polarization
effects).  The  reason is in  necessity to evaluate  the traces of
products of  great number  of Dirac  $\gamma $-matrices.   So  the
problems  arise  in  the  both  cases  of analytical and numerical
calculations   of   the   different   physical   quantities   (for
cumbersomeness of  their expressions).   It  is clear  that direct
calculation of the processes'  amplitudes (see Section II)  is one
of the chances to turn over such a problem.

    In Section III the  possibility to reduce the  expressions for
the traces of products of ten and more Dirac $\gamma $-matrices is
discussed.

Some details can be found in the papers \cite{r1.1} -- \cite{r1.3}.

\section {The method of covariant calculation of the amplitudes of
processes with the polarized Dirac particles}

    There  were  a  lot  of  attempts  to calculate the amplitudes
covariantly  (see, for instance, \cite  {r2.1} -- \cite  {r2.11}).
However, expressions that have been obtained there are  unsuitable
for calculations with interfering diagrams.  The scheme  extending
the mentioned results will be presented below.  It is  concretized
so as one needs in order to avoid the problems with interference.

    There is even number $(2N)$  of fermions in initial and  final state
for any reaction with Dirac particles.  Therefore every diagram contains
$N$ nonclosed fermion lines.  The expression
\begin {equation}
\displaystyle
M_{if} = \bar{u}_f Q u_i
\label{e2.1}
\end {equation}
corresponds to every line in amplitude of process, where $u_i, \;
u_f$ are  Dirac bispinors  for free particles  (for definiteness,  we
anticipate that fermions  are particles.  However, obtained results are
true  in  case  of both  fermions  are antiparticles or one fermion is
particle and another fermion is antiparticle).

$$
\displaystyle
\bar{u} = u^{+} {\gamma}^0 \;\;.
$$

    $Q$ is matrix operator which characterize interaction.  Operator $Q$
is  expressed  as  linear  combination  of  products  of  Dirac  $\gamma
$-matrices (or  of its  contractions with  4-vectors) and  can have  any
number of free Lorentz indexes.

For calculating $M_{if}$ we use the following scheme:
\begin {equation}
\begin {array}{l} \displaystyle
M_{if} = \bar{u}_f Q u_i = ( \bar{u}_f Q u_i ) \cdot
       { \bar{u}_i Z u_f \over \bar{u}_i Z u_f }
   = { Tr ( Q u_i \bar{u}_i Z u_f \bar{u}_f ) \over
       \bar{u}_i Z u_f }
            \\[0.5cm] \displaystyle
\simeq { Tr ( Q u_i \bar{u}_i Z u_f \bar{u}_f ) \over
         | \bar{u}_i Z u_f | }
   = { Tr ( Q u_i \bar{u}_i Z u_f \bar{u}_f ) \over
     [ Tr ( \bar{Z} u_i \bar{u}_i Z u_f \bar{u}_f ) ]^{1/2} }
   = {\cal M}_{if}
\end {array}
\label{e2.2}
\end {equation}
where $Z$   is arbitrary $4 \times 4$ matrix ,
$$
\displaystyle
\bar{Z} = {\gamma}^0 Z^{+} {\gamma}^0
$$
(the symbol $\simeq$  stands for "an equality to within a phase
factor sign").

The projection operators are substituted for $u\bar{u}$ in
(\ref{e2.2}). For particle with mass $m$:
\begin {equation}
\displaystyle
u(p,n) \bar{u}(p,n) = { 1 \over 4m }( \hat{p} + m )
              ( 1 + {\gamma}_5 \hat{n} ) = {\cal P}
\label{e2.3}
\end {equation}
where
$
\displaystyle
\hat{p} = {\gamma}_{\mu} p^{\mu}, \;\; p^2 = m^2, \;\;
n^2 = -1, \;\; pn = 0, \;\; \bar{u} u = 1, \;\;
{\gamma}_5 = {\it i} {\gamma}^0 {\gamma}^1 {\gamma}^2 {\gamma}^3 \;\;.
$
\\
(We use the same metric as in the book \cite{r2.12}:
$$
\displaystyle
a^{\mu} = ( a_0, \vec{a} ),\;\;\;
a_{\mu} = ( a_0, -\vec{a} ),\;\;\;
ab = a_{\mu} b^{\mu} = a_0 b_0 - \vec{a} \vec{b} \, . \, )
$$

For massless particle projection operator is following:
\begin {equation}
\displaystyle
u_{\pm}(q) \bar{u}_{\pm}(q)
= { 1 \over 2 }( 1 \pm {\gamma}_5 ) \hat{q}
 = {\cal P}_{\pm }
\label{e2.4}
\end {equation}
where
$
\displaystyle
q^2 = 0, \;\; \bar{u}_{\pm} {\gamma}_{\mu} u_{\pm} = 2 q_{\mu}
$
(signs $\pm $ correspond to helicity of particle).

In the articles \cite{r2.1}, \cite{r2.2}, \cite{r2.4} one chooses
$$
\displaystyle
Z = 1 \;\;.
$$
    The results of article  \cite{r2.3} come to the  same approach
for  processes  with  massless  particles.    In  \cite{r2.2}  one
proposes
$$
\displaystyle
Z = {\gamma}_5
$$
    too.   The results  of article  \cite{r2.5} come  to the  same
choice.     The  results   of  articles  \cite{r2.6},  \cite{r2.7}
correspond to choice
$$
\displaystyle
Z = 1 + {\gamma}^0 \;\;.
$$
The results of \cite{r2.8}, \cite{r2.9} come to
$$
\displaystyle
Z = 1 + \hat{r}
$$
    (where $r$  is arbitrary  4-momentum, such  as $r^2 = 1$). In
these papers for 4-vectors, which determine axes of spin
projections, one uses
$$
\displaystyle
n_i = { m_i^2 p_f - ( p_i p_f ) p_i \over
        m_i [ ( p_i p_f )^2 - m_i^2 m_f^2 ]^{1/2} } \;\;\; ,
\;\;\;\;
n_f = - { m_f^2 p_i - ( p_i p_f ) p_f \over
        m_f [ ( p_i p_f )^2 - m_i^2 m_f^2 ]^{1/2} } \;\;.
$$
    The results of article  \cite{r2.10} come to
$$
\displaystyle
Z = \hat{k_1} \hat{k_2}
$$
(where $k_1$, $k_2$  are arbitrary  4-vectors, such  as
$k_1^2 = k_2^2 = 0$)
for  processes  with  massless  particles.
    The results of article  \cite{r2.11} come to
$$
\displaystyle
Z=\hat{n}\hat{p}
$$
(where $n$, $p$  are arbitrary  4-vectors, such  as
$n^2 = -1  \;\;,\;\;  p^2 = 0  \; , \;\; (pn) = 0 \;\; $)
for  processes  with  massless  particles.

    However  all  expressions  for   amplitudes  [as  it  follows   from
(\ref{e2.2})] are known to  within a phase factor:
\begin {equation}
\displaystyle
{\cal M}_{if} = M_{if} \cdot { \bar{u}_i Z u_f \over
                             | \bar{u}_i Z u_f | } \;\; .
\label{e2.5}
\end {equation}

    It is obviously that this circumstance creates no problems, when  we
calculate amplitude for alone diagram.  Really
\begin {equation}
\begin {array}{l} \displaystyle
( {\cal M}_{if} )^{*}
= { [ Tr ( Q u_i \bar{u}_i Z u_f \bar{u}_f ) ]^{*}
  \over [ Tr ( \bar{Z} u_i \bar{u}_i Z u_f \bar{u}_f ) ]^{1/2} }
= { [ ( \bar{u}_f Q u_i ) ( \bar{u}_i Z u_f ) ]^{*} \over
     [ Tr ( \bar{Z} u_i \bar{u}_i Z u_f \bar{u}_f ) ]^{1/2} }
           \\[0.5cm] \displaystyle
= { ( \bar{u}_f \bar{Z} u_i ) ( \bar{u}_i \bar{Q} u_f ) \over
  [ Tr ( \bar{Z} u_i \bar{u}_i Z u_f \bar{u}_f ) ]^{1/2} }
= { Tr ( \bar{Z} u_i \bar{u}_i \bar{Q} u_f \bar{u}_f ) \over
  [ Tr ( \bar{Z} u_i \bar{u}_i Z u_f \bar{u}_f ) ]^{1/2} } \;\;.
\end {array}
\label{e2.6}
\end {equation}
$$
\begin {array}{c} \displaystyle
{\cal M}_{if} ({\cal M}_{if} )^{*}
= { Tr ( Q u_i {\bar u}_i Z u_f {\bar u}_f ) \cdot
    Tr ( \bar{Z} u_i {\bar u}_i \bar{Q} u_f {\bar u}_f) \over
    Tr ( \bar{Z} u_i {\bar u}_i Z u_f {\bar u}_f ) }
                        \\[0.5cm] \displaystyle
= { Tr ( Q u_i {\bar u}_i \bar{Q} u_f {\bar u}_f ) \cdot
    Tr ( \bar{Z} u_i {\bar u}_i Z u_f {\bar u}_f ) \over
    Tr ( \bar{Z} u_i {\bar u}_i Z u_f {\bar u}_f ) }
=   Tr ( Q u_i {\bar u}_i \bar{Q} u_f {\bar u}_f )
                        \\[0.5cm] \displaystyle
= ( {\bar u}_f Q u_i ) ( {\bar u}_i \bar{Q} u_f )
= ( {\bar u}_f Q u_i ) ( {\bar u}_f Q u_i )^{*}
= M_{if} ( M_{if} )^{*} = | M_{if} |^2 \;\; .
\end {array}
$$

For calculating $ | M_{if} |^2 $ we used identity
\begin {equation}
\begin {array}{c} \displaystyle
 Tr ( A u_i {\bar u}_i B u_f {\bar u}_f ) \cdot
 Tr ( C u_i {\bar u}_i D u_f {\bar u}_f )
= ( {\bar u}_f A u_i ) ( {\bar u}_i B u_f )
  ( {\bar u}_f C u_i ) ( {\bar u}_i D u_f )
                        \\[0.5cm] \displaystyle
\equiv ( {\bar u}_f A u_i ) ( {\bar u}_i D u_f )
       ( {\bar u}_f C u_i ) ( {\bar u}_i B u_f )
= Tr ( A u_i {\bar u}_i D u_f {\bar u}_f ) \cdot
  Tr ( C u_i {\bar u}_i B u_f {\bar u}_f )
\end {array}
\label{e2.7}
\end {equation}
where $A$, $B$, $C$, $D$ are arbitrary $4 \times 4$-matrices.

    However, in general case, presence of unknown phase factor does  not
enable formula (\ref{e2.2}) to be used for calculation of  amplitudes of
processes  which  proceed  in  a  few  channels  since  expressions  for
amplitudes  which  correspond  to  different  channels are multiplied by
different phase factors. Besides that, ambiguity of the type
$
\displaystyle
{ 0 \over 0 }
$
can appear during the calculations.

But there are no such difficulties if we choose   \\
\hspace*{30mm}
$ Z = {\cal P} \;\;\; $ [see (\ref{e2.3})] $\;\;$
\hspace{2mm} or \hspace{5mm}
$Z = {\cal P}_{\pm} \;\;\; $ [see (\ref{e2.4})] .  \\
Really, in this case the unknown phase factor for each line of
diagram splits into two parts [see (\ref{e2.5})]:
$$
\displaystyle
{\cal M}_{if} = M_{if} \cdot {\bar{u}_i u \bar{u} u_f \over
                             |\bar{u}_i u \bar{u} u_f | }
= M_{if} \cdot { \bar{u}_i u \over | \bar{u}_i u | }
         \cdot { \bar{u} u_f \over | \bar{u} u_f | } \;\; .
\label{e2.8}
$$
    Let us  consider  the diagrams of the process,
which proceeds in 2 different channels in  general form
(see Fig.\ref{Fg1}):

\begin{figure}[ht]
\begin{tabular}{cc}
\begin{picture}(200,100)
\put(100,50){\oval(60,80)}
\put(71,76){\line(1,1){13.}}
\put(70,70){\line(1,1){20.}}
\put(70,65){\line(1,1){25.}}
\put(70,60){\line(1,1){30.}}
\put(70,55){\line(1,1){35.}}
\put(70,50){\line(1,1){40.}}
\put(70,45){\line(1,1){44.}}
\put(70,40){\line(1,1){48.}}
\put(70,35){\line(1,1){51.}}
\put(70,30){\line(1,1){54.}}
\put(71,26){\line(1,1){55.}}
\put(72,22){\line(1,1){56.}}
\put(74,19){\line(1,1){55.}}
\put(76,16){\line(1,1){54.3}}
\put(79,14){\line(1,1){51.}}
\put(82,12){\line(1,1){48.}}
\put(86,11){\line(1,1){44.}}
\put(90,10){\line(1,1){40.}}
\put(95,10){\line(1,1){35.}}
\put(100,10){\line(1,1){30.}}
\put(105,10){\line(1,1){25.}}
\put(110,10){\line(1,1){20.}}
\put(116,11){\line(1,1){13.}}
\put(05,30){\makebox(0,0){$ 2 $}}
\put(05,70){\makebox(0,0){$ 1 $}}
\put(195,30){\makebox(0,0){$ 4 $}}
\put(195,70){\makebox(0,0){$ 3 $}}
\thicklines
\put(10,30){\line(2,0){60.}}
\thicklines
\put(10,70){\line(2,0){60.}}
\thicklines
\put(130,30){\line(2,0){60.}}
\thicklines
\put(130,70){\line(2,0){60.}}
\put(40,30){\vector(1,0){0.}}
\put(40,70){\vector(1,0){0.}}
\put(160,30){\vector(1,0){0.}}
\put(160,70){\vector(1,0){0.}}
\end{picture}
&
\begin{picture}(200,100)
\put(100,50){\oval(60,80)}
\put(71,76){\line(1,1){13.}}
\put(70,70){\line(1,1){20.}}
\put(70,65){\line(1,1){25.}}
\put(70,60){\line(1,1){30.}}
\put(70,55){\line(1,1){35.}}
\put(70,50){\line(1,1){40.}}
\put(70,45){\line(1,1){44.}}
\put(70,40){\line(1,1){48.}}
\put(70,35){\line(1,1){51.}}
\put(70,30){\line(1,1){54.}}
\put(71,26){\line(1,1){55.}}
\put(72,22){\line(1,1){56.}}
\put(74,19){\line(1,1){55.}}
\put(76,16){\line(1,1){54.3}}
\put(79,14){\line(1,1){51.}}
\put(82,12){\line(1,1){48.}}
\put(86,11){\line(1,1){44.}}
\put(90,10){\line(1,1){40.}}
\put(95,10){\line(1,1){35.}}
\put(100,10){\line(1,1){30.}}
\put(105,10){\line(1,1){25.}}
\put(110,10){\line(1,1){20.}}
\put(116,11){\line(1,1){13.}}
\put(05,30){\makebox(0,0){$ 2 $}}
\put(05,70){\makebox(0,0){$ 1 $}}
\put(195,30){\makebox(0,0){$ 3 $}}
\put(195,70){\makebox(0,0){$ 4 $}}
\thicklines
\put(10,30){\line(2,0){60.}}
\thicklines
\put(10,70){\line(2,0){60.}}
\thicklines
\put(130,30){\line(2,0){60.}}
\thicklines
\put(130,70){\line(2,0){60.}}
\put(40,30){\vector(1,0){0.}}
\put(40,70){\vector(1,0){0.}}
\put(160,30){\vector(1,0){0.}}
\put(160,70){\vector(1,0){0.}}
\end{picture}
\end{tabular}
\caption{ The diagrams of the process, which proceeds in 2 different
channels in general form.}
\label{Fg1}
\end{figure}
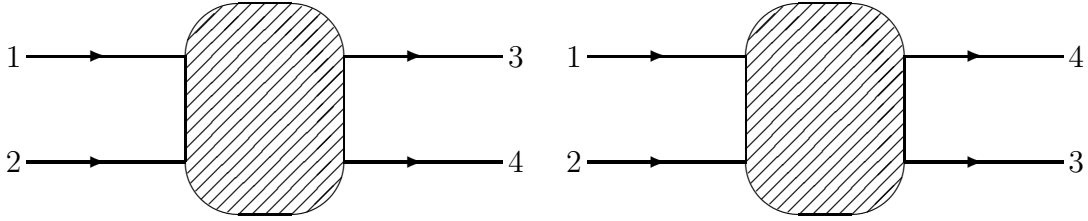

The expression
$$
\displaystyle
M = ( \bar{u}_3 Q u_1 ) \cdot ( \bar{u}_4 R u_2 )
= M_{13} \cdot M_{24}
$$
corresponds to the first diagram. The expression
$$
\displaystyle
M' = ( \bar{u}_4 S u_1 ) \cdot ( \bar{u}_3 T u_2 )
            = M_{14} \cdot M_{23}
$$
corresponds to  the second  diagram, where  $Q$, $R$, $S$, $T$
are arbitrary matrix operators which characterize interaction.

Calculating the amplitudes we have
$$
\begin {array}{c} \displaystyle
{\cal M} = M_{13} \cdot M_{24}
          \cdot { \bar{u}_1 u \over | \bar{u}_1 u | }
          \cdot { \bar{u} u_3   \over  | \bar{u} u_3 | }
          \cdot { \bar{u}_2 u  \over  | \bar{u}_2 u | }
          \cdot { \bar{u} u_4  \over  | \bar{u} u_4 | } \;\; ,
                        \\[0.5cm] \displaystyle
{\cal M}' = M_{14} \cdot M_{23}
        \cdot { \bar{u}_1 u  \over  | \bar{u}_1 u | }
        \cdot { \bar{u} u_4  \over  | \bar{u} u_4 | }
        \cdot { \bar{u}_2 u  \over  | \bar{u}_2 u | }
        \cdot { \bar{u} u_3  \over  | \bar{u} u_3 | } \;\; ,
\end {array}
$$
that is the amplitudes of the different diagrams have the same phase
factor.

    Thus  we  must  calculate  amplitude  of  process  with  interfering
diagrams in form
\begin {equation}
\displaystyle
{\cal M} + {\cal M}'
= { Tr ( Q {\cal P}_1 {\cal P} {\cal P}_3 ) \cdot
    Tr ( R {\cal P}_2 {\cal P} {\cal P}_4 )
  + Tr ( S {\cal P}_1 {\cal P} {\cal P}_4 ) \cdot
    Tr ( T {\cal P}_2 {\cal P} {\cal P}_3 ) \over
  [ Tr ( {\cal P} {\cal P}_1 ) Tr ( {\cal P} {\cal P}_2 )
    Tr ( {\cal P} {\cal P}_3 )
    Tr ( {\cal P} {\cal P}_4 ) ]^{1/2} } \;\;.
\label{e2.9}
\end {equation}
    This  expression  enable  to  calculate  the  amplitude numerically.
Complex  numbers   being  obtained   under  calculation   are  used  for
calculation of cross section of process.

    For the denominator in the right-hand side of (\ref{e2.9}) we
take into account the identity
$$
\displaystyle
Tr ( \bar{\cal P} {\cal P}_i {\cal P} {\cal P}_f ) =
Tr ( {\cal P} {\cal P}_i ) Tr ( {\cal P} {\cal P}_f ) \;\;,
$$
since projection operators have the following properties
$$
\displaystyle
\bar{\cal P} = {\cal P} \;\; , \;\;\;\;\;
{\cal P} A {\cal P}
= Tr [ {\cal P} A ] \cdot {\cal P} \;\; ,
$$
$$
\displaystyle
\bar{\cal P}_{\pm} = {\cal P}_{\pm} \;\; , \;\;\;\;\;
{\cal P}_{\pm} A {\cal P}_{\pm}
= Tr [ {\cal P}_{\pm} A ] \cdot {\cal P}_{\pm}
\;\;\; .
$$
Really
$$
\begin {array}{l} \displaystyle
\bar{\cal P} = {\gamma}^0 {\cal P}^{+} {\gamma}^0
= {\gamma}^0 ( u \bar{u} )^{+} {\gamma}^0
= {\gamma}^0 ( u u^{+} {\gamma}^0 )^{+} {\gamma}^0
                   \\[0.5cm] \displaystyle
= {\gamma}^0 [ ( {\gamma}^0 )^{+} ( u^{+ } )^{+} u^{+} ]
  {\gamma}^0
= {\gamma}^0 [ {\gamma}^0 u u^{+} ] {\gamma}^0
= u u^{+} {\gamma}^0
= u\bar{u}={\cal{P} } \;\; ,
\end {array}
$$
$$
\begin {array}{l} \displaystyle
{\cal P} A {\cal P} = (u)_{\alpha} ( \bar{u} )_{\beta}
(A)^{\beta \rho} (u)_{\rho} ( \bar{u} )_{\delta}
= [ ( \bar{u} )_{\beta} (A)^{\beta \rho} (u)_{\rho} ]
    (u)_{\alpha} ( \bar{u} )_{\delta}
                   \\[0.5cm] \displaystyle
= [ (u)_{\rho} ( \bar{u} )_{\beta} (A)^{\beta \rho} ]
  (u)_{\alpha} ( \bar{u} )_{\delta}
= Tr [ {\cal P} A ] \cdot {\cal P} \;\; .
\end {array}
$$

    If in individual cases  under numerical calculations denominator  in
(\ref{e2.9}) is  equal to  $0$, it  is sufficiently  to change arbitrary
projection operator ${\cal P}$, which appears in the expression
(\ref{e2.9}).

    Notice that calculation of amplitude for alone diagram is easer than
calculation  of  squared  amplitude,  if  operators,  which characterize
interaction, contain product of greater number $\gamma$-matrices,  then
projection operator.

    Really, when number of  $\gamma$-matrices in operator  $Q$ increase
by $I$ [see  (\ref{e2.2})], theirs number  in numerator of  (\ref{e2.2})
increase only by $I$ (denominator does not change), but in  construction
$Tr ( Q u_i \bar{u}_i \bar{Q} u_f \bar{u}_f )$ , which appear,
when
we calculate squared  matrix element, the  number of $\gamma  $-matrices
increase by $2I$.   We take into account  that trace of product  of $2J$
$\gamma $-matrices contains $1\cdot  3\cdot 5\cdot \ldots \cdot  (2J-1)$
terms and we obtain  that the more complicated  is a process the  bigger
are advantageous being given under calculation of this process by method
of directly calculation of amplitudes.

    However,  for   processes  with   interfering  diagrams   method  of
calculation of amplitudes  is easier in  any  case, because  we need not
calculate the interference terms.

    As it was mentioned before, it is simply to make a generalization of
this method for a reactions with participation of antiparticles.  It  is
sufficiently for it to substitute projection operators of  antiparticles
in place of operators of particles.  Let, for example, we are interested
in value $ \bar{v}_f Q  u_i$, where $v_f$ is bispinor for free
antiparticle.  Then
\begin {equation}
\displaystyle
\bar{v}_f Q u_i = { Tr ( Q u_i \bar{u}_i Z v_f \bar{v}_f )
\over [ Tr ( \bar{Z} u_i \bar{u}_i Z v_f \bar{v}_f ) ]^{1/2} }
\label{e2.10}
\end {equation}
where
$$
\displaystyle
v(p,n) \bar{v}(p,n) = { 1 \over 4m } ( -m + \hat{p} )
( 1 + {\gamma}_5 \hat{n} )
$$
for massive antiparticle, or
$$
\displaystyle
v_{\pm}(q) \bar{v}_{\pm}(q)
= { 1 \over 2 }( 1 \mp {\gamma}_5 ) \hat{q}
$$
for massless antiparticle.
As always, we use  (\ref{e2.3}) or (\ref{e2.4}) instead of $Z$ .

    Notice that in (\ref{e2.10})  and in further consideration  we
shall use equality sign instead of symbol $\simeq $ , since  there
exist not any trouble with phase factors already.

    As an example, we give formulas for  calculation of amplitudes of
processes with massless Dirac particles. In this case formula
(\ref{e2.2}) takes the next form
\begin {equation}
\displaystyle
\bar{u}_{\pm}(p_3) Q u_{\pm}(p_1)
= { Tr [ Q \hat{p}_1 \hat{q} \hat{p}_3 ( 1 \mp {\gamma}_5 ) ]
\over 4 [ ( q p_1 ) ( q p_3 ) ]^{1/2} } \;\;\;.
\label{e2.11}
\end {equation}
Here
$
\displaystyle
Z = { 1 \over 2 } ( 1 \mp {\gamma}_5 ) \hat{q}
= {\cal P}_{\mp} \;\; , \;\;\;
q^2 = 0.
$
\\
    Massless 4-vector $q$ can be arbitrary, but it must be the same  for
all considered nonclosed fermion lines of diagrams.

\begin {equation}
\displaystyle
\bar{u}_{\pm}(p_3) Q u_{\mp}(p_1) = { Tr [ Q \hat{p}_1
( m \pm \hat{n} \hat{p} ) \hat{p}_3 ( 1 \mp {\gamma}_5 ) ] \over
4 \{ [ ( p p_1 ) \pm  m ( n p_1 ) ]
     [ ( p p_3 ) \mp m ( n p_3 ) ] \} ^{1/2} }
\;\;\; .
\label{e2.12}
\end {equation}
Here
$
\displaystyle
Z = { 1 \over 4m } ( m + \hat{p} )
( 1 + {\gamma}_5 \hat{n} ) = {\cal P}, \;\;\;
p^2 =  m^2, \;\;\; n^2 = -1,  \;\;\; pn = 0.
$
\\
    As regards 4-vectors $p$ and $n$ the same observation as the one for
vector $q$  in (\ref{e2.11}) is right.

    In the last  case we can  not use easier  operator ${\cal{P}}_{\pm}$
for $Z$, since in this case numerator and denominator are identical with
0. But  we  may to  require for  maximum  simplicity of calculations
$m = 0, \;\;\; p^2 = 0 \;\;$ in (\ref{e2.12}):
\begin {equation}
\displaystyle
\bar{u}_{\pm }(p_3) Q u_{\mp}(p_1) = \pm { Tr [ Q \hat{p}_1
 \hat{n} \hat{p} \hat{p}_3 ( 1 \mp {\gamma}_5 ) ] \over
4 [ ( p p_1 ) ( p p_3 ) ]^{1/2} }
\;\;\; .
\label{e2.13}
\end {equation}
    Formula (\ref{e2.13}) generalize method of calculation
of amplitudes being offered in \cite{r2.11}.

    If under numerical  calculations denominator in  (\ref{e2.11})
--  (\ref{e2.13})  is  equal  to  0  for  some  values $p_1$ and
$p_3$,  it  is  sufficiently  to  change  values  of   arbitrary
4-vectors  $q$  or  $p$, $n$   being  contained  by  these
formulae
(simultaneously for all lines of diagrams being considered).

\section {Reducing of expressions for the traces of products of ten
and more Dirac $\gamma$-matrices}

Let us consider the determinant
\begin {equation}
\begin {array}{l} \displaystyle
\left| \matrix{
g_{\mu \nu}   & g_{\mu \alpha}  &
g_{\mu \beta} & g_{\mu \lambda} &
g_{\mu \rho}  \\
g_{\sigma \nu}   & g_{\sigma \alpha}  &
g_{\sigma \beta} & g_{\sigma \lambda} &
g_{\sigma \rho}  \\
g_{\tau \nu}   & g_{\tau \alpha}  &
g_{\tau \beta} & g_{\tau \lambda} &
g_{\tau \rho}  \\
g_{\kappa \nu}   & g_{\kappa \alpha}  &
g_{\kappa \beta} & g_{\kappa \lambda} &
g_{\kappa \rho}  \\
g_{\omega \nu}   & g_{\omega \alpha}  &
g_{\omega \beta} & g_{\omega \lambda} &
g_{\omega \rho}
} \right| \equiv 0 \;\;
\end {array}
\label{e3.1}
\end {equation}
where
$$
\displaystyle
g_{\mu \nu} =
\left\{
\begin{array}{rl}
               1 & $if$\;\;\; \mu = \nu = 0  \\
              -1 & $if$\;\;\; \mu = \nu = 1,2,3  \\
               0 & $if$\;\;\; \mu \neq \nu
\end{array}
\right.
$$
The validity of this identity follows from the fact that the
expression  on the left-hand side of (\ref{e3.1}) is completely
antisymmetric with respect to each of five indices:
${\nu}$, ${\alpha}$, ${\beta}$, ${\lambda}$, ${\rho}$
(and ${\mu}$, ${\sigma}$, ${\tau}$, ${\kappa}$, ${\omega}$ too).
In four-dimensional space, every tensor that is antisymmetric with
respect to more than four indices vanishes identically, since
the values of at least two of them must be equal.

By analogy with the Gram determinant, we introduce the notation
(see \cite{r3.1})
$$
\begin {array}{l} \displaystyle
\left| \matrix{
g_{\mu \nu} & g_{\mu \alpha} & g_{\mu \beta} & g_{\mu \lambda} &
g_{\mu \rho}  \\
g_{\sigma \nu} & g_{\sigma \alpha} & g_{\sigma \beta} &
g_{\sigma\lambda} & g_{\sigma \rho}  \\
g_{\tau \nu} & g_{\tau \alpha} & g_{\tau \beta} & g_{\tau \lambda} &
g_{\tau \rho}  \\
g_{\kappa \nu} & g_{\kappa \alpha} & g_{\kappa \beta} & g_{\kappa \lambda} &
g_{\kappa \rho}  \\
g_{\omega \nu} & g_{\omega \alpha} & g_{\omega \beta} & g_{\omega \lambda} &
g_{\omega \rho}
} \right| =
G\pmatrix{\mu & \sigma & \tau  & \kappa  & \omega \\
          \nu & \alpha & \beta & \lambda & \rho }  \;\; .
\end {array}
$$

It follows from the properties of determinants that
\begin {equation}
\begin {array}{l} \displaystyle
G\pmatrix{\mu & \sigma & \tau  & \kappa  & \omega & \chi \\
          \nu & \alpha & \beta & \lambda & \rho   & \varphi}
\equiv 0 \;\; ,
\end {array}
\label{e3.2}
\end {equation}

\begin {equation}
\begin {array}{l} \displaystyle
G\pmatrix{\mu & \sigma & \tau  & \kappa  & \omega & \chi    & \xi \\
          \nu & \alpha & \beta & \lambda & \rho   & \varphi & \pi}
          \equiv 0
\end {array}
\label{e3.3}
\end {equation}
etc. We note that the identity (\ref{e3.1}) is valid only in a
space of dimension not higher than 4, the identity
(\ref{e3.2}) is valid in a space of dimension not higher than 5,
etc.

   Identities of the type (\ref{e3.1}), (\ref{e3.2}), (\ref{e3.3})
and others like them can be used to simplify the expressions for
the traces of a product  of 10 and more Dirac ${\gamma}$-matrices.
For example, the expression for
$$
\displaystyle
Tr( {\gamma}_{\mu}
    {\gamma}_{\sigma}
    {\gamma}_{\tau}
    {\gamma}_{\kappa}
    {\gamma}_{\omega}
    {\gamma}_{\nu}
    {\gamma}_{\alpha}
    {\gamma}_{\beta}
    {\gamma}_{\lambda}
    {\gamma}_{\rho} )
$$
calculated by means of the computer system  REDUCE, contains
945 terms. However, the expression
$$
\begin {array}{c} \displaystyle
Tr( {\gamma}_{\mu}
    {\gamma}_{\sigma}
    {\gamma}_{\tau}
    {\gamma}_{\kappa}
    {\gamma}_{\omega}
    {\gamma}_{\nu}
    {\gamma}_{\alpha}
    {\gamma}_{\beta}
    {\gamma}_{\lambda}
    {\gamma}_{\rho} )
- 4 \cdot G\pmatrix{\mu & \sigma & \tau  & \kappa  & \omega \\
                    \nu & \alpha & \beta & \lambda & \rho }
- 4 \cdot G\pmatrix{\mu    & \sigma & \tau   & \kappa & \rho \\
                    \omega & \nu    & \alpha & \beta  & \lambda }
            \\[0.5cm] \displaystyle
- 4 \cdot G\pmatrix{\mu    & \sigma & \tau  & \lambda & \rho \\
                    \kappa & \omega & \nu   & \alpha  & \beta }
- 4 \cdot G\pmatrix{\mu  & \sigma & \beta  & \lambda & \rho \\
                    \tau & \kappa & \omega & \nu     & \alpha }
- 4 \cdot G\pmatrix{\mu    & \alpha & \beta  & \lambda & \rho \\
                    \sigma & \tau   & \kappa & \omega  & \nu  }
            \\[0.5cm] \displaystyle
+ 4 \cdot G\pmatrix{\mu    & \tau   & \omega & \alpha  & \lambda \\
                    \sigma & \kappa & \nu    & \beta   & \rho }
+ 4 \cdot G\pmatrix{\mu  & \sigma & \kappa & \omega  & \beta \\
                    \tau & \nu    & \alpha & \lambda & \rho }
+ 4 \cdot G\pmatrix{\mu    & \tau   & \kappa & \alpha & \rho \\
                    \sigma & \omega & \nu    & \beta  & \lambda }
\end {array}
$$
which is identical to it, contains only 531 terms.


\end {document}